# IoT Based Smart Home Using Blynk Framework

Bharat Bohara* and Sunil Maharjan†

*Department of Electrical and Electronics Engineering

†Department of Mechanical Engineering

Kathmandu University, Dhulikhel, Nepal

* email: bharatbohara@hotmail.com

Bibek Raj Shrestha

Department of Electronics and Computer Engineering

IoE, Pulchowk, Lalitpur, Nepal

email: bbkstha@gmail.com

*Abstract*—The project discussed in this paper is targeted at solving sundry problems faced by Nepalese people in their daily life. It is designed to control and monitor appliances via smartphone using Wi-Fi as communication protocol and raspberry pi as private server. All the appliances and sensors are connected to the internet via NodeMcu microcontroller, which serves as the gateway to the internet. Even if the user goes offline, the system is designed to switch to automated state controlling the appliances automatically as per the sensors readings. Also, the data are logged on to the server for future data mining. The core system of this project is adopted from the Blynk framework**.

## I. INTRODUCTION

Today, internet has become an integral part of people's lives, influencing the daily activities of almost every human being. Evidently, every second smart phones with sophisticated functionalities are released out in the market. It infers that internet users in accordance with the booming smartphone use are multiplying vigorously day by day. Thus, connecting every- thing possessed by a human to the internet and subsequently monitoring and further controlling through smartphones is the ultimate goal of this project.

IoT is the area of network in connection with consequences, result and actions via internet allowing them to send and receive data [2]. Here, things are connected among themselves without human intervening for automatic identification of intended activities. IoT helps in sharing of information from sensors through wireless network, achieving identification and informational exchange in open computing network and achieving transparent management of system. Things that we are using in our daily life are becoming smart with the current technologies but it isnt enough until we link them to act with the changing environment and additionally make their own inter-network, that is, machine-to-machine communication.[2] In a dynamically changing city areas, creating and maintaining public transport system, smart provision of electric energy, water and gas distribution systems, waste management and maintenance of the city infrastructure like roads and public parks are some of the challenging activities to be taken care of. We believe these complex systems will be better addressed with IoT technology.

### A. Problem Statement and Significance

In Nepal, load shedding has been an incorrigible and impasse problem for more than a decade now. During dry seasons, electricity cut-off reaches up to 16 hours a day [3]. Hence, people are obliged to make troublesome decisions to comply with the power-cut schedule. In cities, most of the people are job-indulged and are busy at the day time, which, unfortunately, is low peak-load hour when electricity is available at home. On the other hand, at the time when people reach their home, everybody turn ON their appliances that makes peak load hour, where load-shedding is scheduled to maximum extent. So, there is a need of a system that can automate household tasks such as filling water tanks, charging devices and such, in the absence of the house owners.

Many incidents of robbery are witnessed in cities such as Kathmandu, Pokhara, Biratnagar, etc. Houses, shops, offices, industries, business complexes, etc. are no longer safe. Ill-minded people are always in search of right opportunity to rob these places. Thus, information regarding intruder detection and alarms are needed for security. Above all, security is critical and also is the foremost priority of this research project. Most of acreage land in Nepal is situated in rural areas where there is shortage of manpower owing to migration of young and energetic people to either cities or abroad for jobs. The remaining people in rural areas are mostly children and old-aged persons who are unable to look after their farm and cultivation. On such circumstances, Nepal is moving towards the phase of full dependence even on agricultural imports, let alone the massive imports of industrial products. By introducing smart farming and agriculture [4], need of capable workers in those remote places can be compensated and agricultural products can be grown with less manpower.

Similarly, traffic problems can be frustrating as evident from the hours of traffic jams and pollution in Kathmandu. Rapid influx of people in city areas and outrage increment of vehicles mostly two-wheelers and congestion of road are the major root causes of inevitable daily traffic jam. This project can assist traffic management using the smart traffic management system. Beyond above mentioned problems in Nepal, there are many other arenas where IoT technology can be beneficially applied.

---
**Blynk [1] is an open source IoT platform for IoT technology. http://www.blynk.cc/

## II. LITERATURE REVIEW

Smart cities based on IoT technology are becoming more and more popular. Initially, IoT's goal was to connect physical devices to internet. Then, Web of Things (WoT) emerged to easily connect sensors to the web, get the data and exchange data on the web that has been produced by the devices [5]. We have gone thoroughly through number of journals, research and conference papers and project reports to thoroughly understand the concept of IoT technology. Similarly, we have researched various IoT based projects that have been designed and de- veloped in the past. Some of the proposed and existing smart cities platforms are as follows.

The READY4SmartCities [6] aims at reducing energy consumption and CO2 emission in cities exploiting ontologies and linked data. This project is intended to generate and provide energy-related data such as climatic, pollution, traffic, activity etc. But it doesnt encompass vital IoT domains like healthcare, smart farm etc. and neither does it mention need to integrate a reasoning engine to analyze IoT data.

The STAR-CITY project is deployed in four cities: Dublin, Bologna, Miami and Rio [7]. They use semantic web technologies to diagnose and predict road traffic congestions. As per their design, they use six heterogeneous sources: road weather conditions, weather information, Dublin bus stream, social media feeds, road works and maintenance, and city events. They use Semantic Web Rule Language (SWRL) rules such as heavy traffic flow. The project is mainly focused on the traffic analysis.

The CityPulse [8] project is designed for public parking space availability prediction, real time travel planner, air pollution counter-measures, and opting efficient rout and public transport. The project is focused on large-scale analysis and real-time processing.

The SmartSantander[9] project deployed 20,000 sensors measuring temperature, humidity, particles, CO and NO2 for monitoring parks and gardens irrigation, outdoor parking area management, traffic intensity monitoring, and smart metering. The study of above mentioned projects have helped us design and develop a prototypic system that could best address the relevant issues prevalent in our country Nepal.

## III. IOT ARCHITECTURE

The physical layer consists of the devices that are to be controlled. The sensors to sense the surrounding environmental conditions are also connected to this layer. The data link layer consists of IoT gateway router (here, we have used NodeMcu as router gateway), device manager and various communication protocols. This layer links the home appliances to the webserver or cloud via Wi-Fi communication. Raspberry pi is used as private server to store the sensor data and also it sends data to the end users upon request. In this system, raspberry pi falls under database/server layer. The application and presentation layer consist of web protocol. This layer con- stitute either designing of a webpage for accessing the devices connected to the perception layer via PC or laptop computer, or building an android or iOS mobile application if the devices are to be controlled and monitored via smartphones. The layers of IoT for the proposed system are shown below:

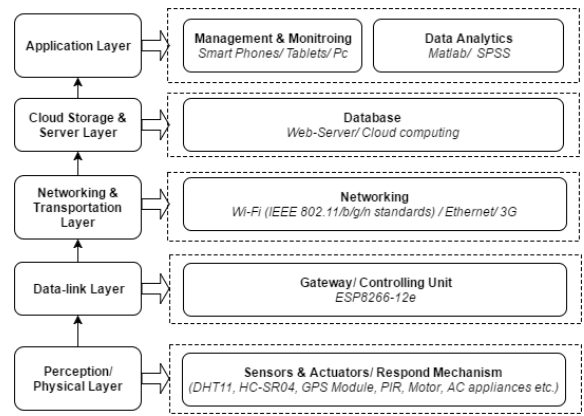

Fig. 1: IoT Architecture of the Project

## IV. HARDWARE DESIGN

The system is divided into two major parts: software and hardware design. Hardware configuration involves arranging microprocessor, microcontroller, sensors and actuators whereas software portion encloses programming that is written and uploaded in each of the microcontrollers and microprocessor. The system consists of microcontroller connected to sensors and electrical devices that are to be monitored and controlled. This section shows how different hardware components are set up. The specifications and information regarding various components used in this system are descriptively explicated below.

### A. Microprocessor Unit (CPU) Raspberry pi

The Raspberry Pi is a series of credit card-sized single-board computers developed in the United Kingdom by the Raspberry Pi Foundation with the intent to promote the teaching of basic computer science in schools and developing countries.

Raspberry pi 2 is a 900MHz quad-core ARM Cortex-A7 CPU with 1GB RAM. It consists of 4 USB ports, 40 GPIO pins, Full HDMI port, Ethernet port, combined 3.5mm audio jack and composite video, camera interface (CSI), Display interface (DSI), micro SD card slot and a video-core IV 3D graphics core. It has an ARMv7 processor and hence it can run the full range of ARM GNU/Linux distributions, including Ubuntu Core, as well as Microsoft Windows 10. SD cards are used to store the operating system and program memory in either the SDHC or MicroSDHC sizes. Lower level output is provided by a number of GPIO pins which support common protocols like I2C.

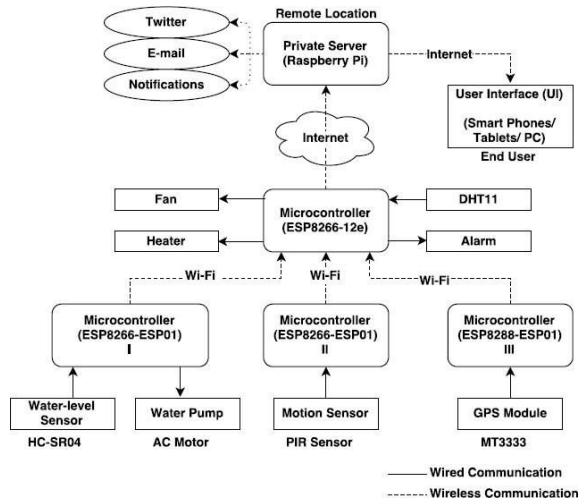

Fig. 2: Schematic Plan of the Project

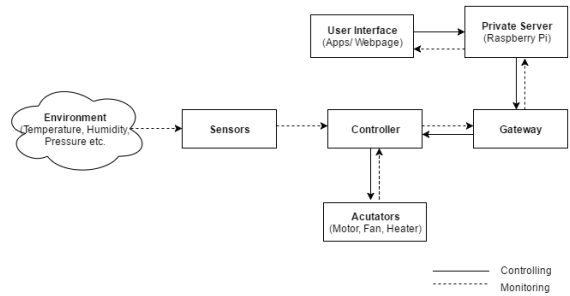

Fig. 3: Working Plan of the Proposed System

## B. Microcontroller Unit (MCU) - ESP8266

The ESP8266 is a low-cost Wi-Fi chip with full TCP/IP stack and microcontroller capability produced by Shanghai-based Chinese manufacturer i.e. Espressif [10]. The chip first came to the attention of western makers in August 2014 with the ESP-01 module, made by a third-party manufacturer - AI-Thinker. This small module allows microcontrollers to connect to a Wi-Fi network and make simple TCP/IP connection using Hayes-style commands.

ESP8266EX requires minimal external circuitry and integrates a 32-bit Tensilica MCU, standard digital peripherals interfaces, PCB traced antennal, RF balun, power amplifier, low noise receive amplifier, filters and power management modules; running at 80MHz. It consists of 64 KiB of instruction RAM, and 96KiB of data RAM. It does have up to 16 GPIO pins and advocates SPI as well as I2C communication networking [10]. It is ultra-low power consuming device with ample processing speed.

## C. Sensor Units

A myriad number of sensors are being used in this system to measure various parameters of environment. The system decisions and operations are controlled by the sensors readings. Ambient temperature and humidity is measured by DHT11 sensor, water-level in tank is incessantly measured by ultrasonic sensor, motion is detected by IR sensor and path/ location of a remote vehicle is tracked by a GPS sensor.

## D. User Interface

The graphical interfaces in smart phones and tablets are designed in the form of android and iOS applications by putting buttons, graph-plotter, LCD and sensor-value display. The user can simply download the app, log in and then monitor and control her entire home appliances. The interface should enable the user to look at the device status and regulate them.

Similarly, UI embodies a web portal for inspecting the status and controlling them from desktop PCs and laptops. Web portal is designed by making webpages from where the user can log in and access the status of her entire home appliances and control them through desktop PCs.

## V. WORKING OF THE SYSTEM

The system consists of three isolated sub-systems: first sub-system consisting of GPS module to get the geo-location, second sub-system consisting of multiple sensors DHT11 tem- perature sensor to measure temperature, PIR sensor to detect motion and ultrasonic sensor to measure the distance, and the third sub- system consisting of a master microcontroller which function as the central coordinator that communicates with other subsystems via Wi-Fi. The master microcontroller is also interfaced with a relay module to control the appliances at the site. The sensor data are fetched to the user interface facilitated by smartphones or tablets from the various sensors using a raspberry pi as the private server.

Basically, control of turning ON or OFF the whole system is at owners hand. As the system gets powered up, it searches for the preset SSID (Service Set Identifier) and connects automatically to the Internet otherwise remains offline and performs the automated-controlling job that doesn't require commands from the owner.

Sensors accumulate disparate ambient-conditions and transmit them to the Microcontroller which processes the data transmitted by each sensor separately and then concurrently send the acquired data to the web server. The readings of each sensor can be accessed by the user from any place at any time. Additionally, all the sensors data are logged per second for future data analysis purpose. Data-logging is done both in microSD card and in the server.

The system operates in two modes automatic mode and manual mode. When it is set at automatic mode, all the home appliances like fan, heater etc. are automated to operate as per the surrounding environmental conditions sensed by the sensors. On the other hand, when it is set at manual mode, the user can locally or remotely monitor and control each of the home appliances via her smart phone or from her office-desktop PC. To recapitulate, the surveillance and control of entire household appliances is under her finger tip.

Beyond auto-manual mode of operation, some of the activities like regular water level (in water-storage tank) inspection and turning ON or OFF the water pump, ringing alarms when detecting human-presence at the door which can also be used for security by installing the motion detectors on home or office so as to alert owner remotely. The user doesnt need to switch these devices at regular interval as they are self- run under the commands of the microcontroller continuously round the clock.

Similarly, some of the alert signals and warnings like house on fire, gas leakage, intruders detection etc. are wirelessly conveyed and notified about the situation to the house owner via e-mail and phone notification. Using GPS, the owner can track and locate her car at any moment from anywhere. To sum up, IoT provides greater extent of security- physical as well psychological.

## VI.  RESULTS AND APPLICATIONS

The proposed prototype of the system was designed and exhibited on the 13th National Technological Festival, Locus-2016, Pulchowk campus, Nepal.

Three different isolated sub-systems : i) relay module system connected to the home appliances to be controlled ii) GPS module and Temperature sensor connected system iii) PIR sensor for motion detection and ultrasonic senor for measuring the water level in the tank were linked to one another via Wi-Fi using NodeMcu controller chip. For user interface, android version of Blynk app with custom designed layout  and buttons was used to facilitate monitoring and controlling various connected things.

The screenshots of the results of the designed system obtained on android application are shown in Figure 4.

By pressing virtual button on the smartphone, the home appliances can be controlled from any remote location. One advantage of this app is that it can be shared within all the family members of the house. When one member switches ON or OFF an appliance, the action will be apparent to all other members sharing the app. Similarly, real-time as well as historical data of measurements of temperature, humidity, GPS location and distance-measure can be obtained from anywhere using the app.

Further, this system can be employed in many places such as banks, hospitals, laboratories, traffic stations, residential apartments, house, streets, poultry farms, greenhouse etc. In a nutshell, this system can be used at multiple fields and areas in order to make them operate smartly.

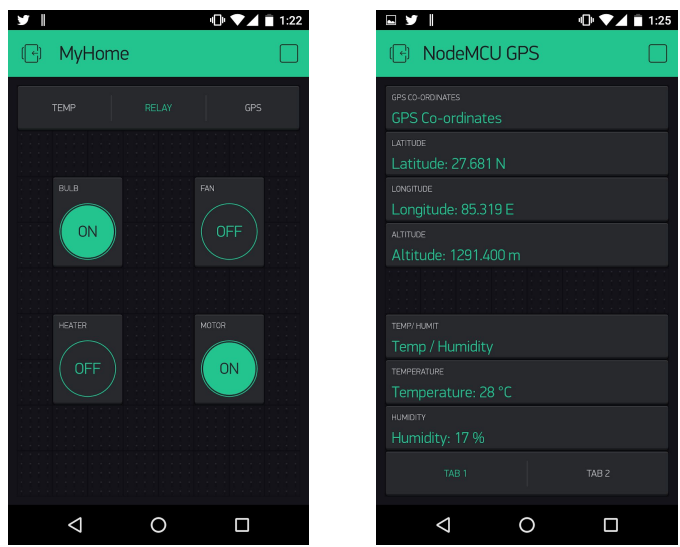

Fig. 4: Screenshots showing appliance switches and sensor outputs

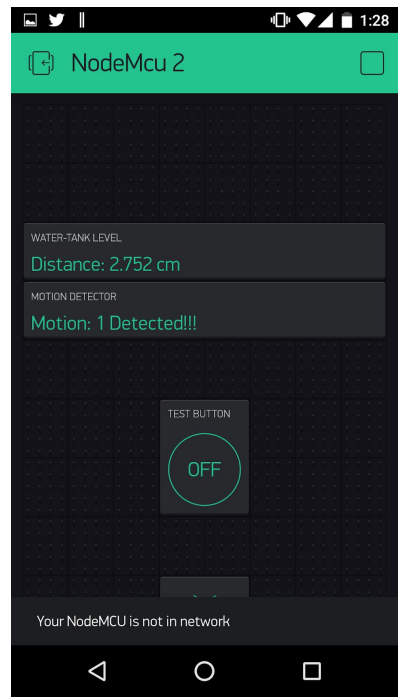

Fig. 5: Screenshots showing distance measured and motion detected.

## VII.  CONCLUSION AND FUTURE WORKS

In this paper, we have introduced a home management and security system. This paper is mainly focused on overcoming everyday problems faced by the people in Nepal where regular power cut-off, unmanaged urbanization, lack of manpower in agriculture and farming, etc are blatantly evident. Our prototypical system is applicable to real-time home security, automation, monitoring and controlling of remote systems.

The future works include: 1) Implementation of the project in one of the remote parts of Nepal to collect data like wind speed, solar irradiance etc. for future research 2) Feasibility study and implementation of the project in an unmanaged and polluted cities like Kathmandu 3) Application of the project in residential complexes and small-scale industries [11] [12] and traditional greenhouses[4] 4) Big data analytics on the collected data using appropriate tools and techniques [13].

## Acknowledgment

We would like to thank Asst. Prof. Dr. Arun K. Timalsina, IOE, Pulchowk Campus, Tribhuvan University for guiding us throughout this research project. We are also very much thankful towards Blynk organization for granting us permission to access its cloud server along with its resources and library files.